\documentclass[conference]{IEEEtran}
\IEEEoverridecommandlockouts
\usepackage{cite}
\usepackage{amsmath,amssymb,amsfonts}
\usepackage{algorithmic}
\usepackage{graphicx}
\usepackage{textcomp}
\usepackage{xcolor}
\usepackage{tabularx}
\usepackage{datatool}
\usepackage{float}

\def\BibTeX{{\rm B\kern-.05em{\sc i\kern-.025em b}\kern-.08em
		T\kern-.1667em\lower.7ex\hbox{E}\kern-.125emX}}

\begin{document}
	\newcolumntype{L}[1]{>{\raggedright\arraybackslash}p{#1}}
	\newcolumntype{C}[1]{>{\centering\arraybackslash}p{#1}}
	\newcolumntype{R}[1]{>{\raggedleft\arraybackslash}p{#1}}
	
	\title{Exploring the Journey to Drug Overdose: Applying the Journey to Crime Framework to Drug Sales Locations and Overdose Death Locations}
	
	\makeatletter
	\newcommand{\linebreakand}{
	\end{@IEEEauthorhalign}
	\hfill\mbox{}\par
	\mbox{}\hfill\begin{@IEEEauthorhalign}
	}

	
	\makeatother
	\author{
		
		\IEEEauthorblockN{Murat Ozer}
		\IEEEauthorblockA{\textit{School of Information Technology} \\
			\textit{University of Cincinnati}\\
			Cincinnati, Ohio, USA \\
			m.ozer@uc.edu}
		\and
		\IEEEauthorblockN{Ismail Onat}
		\IEEEauthorblockA{\textit{Criminal Justice and Criminology Department} \\
			\textit{University of Scranton}\\
			Scranton, PA, USA \\
			ismail.onat@scranton.edu}
		\and
		\IEEEauthorblockN{Halil Akbas}
		\IEEEauthorblockA{\textit{Criminal Justice Department} \\
			\textit{Troy University}\\
			Troy, AL, USA \\
			hakbas@troy.edu}
		\linebreakand 
		\IEEEauthorblockN{Nelly Elsayed}
		\IEEEauthorblockA{\textit{School of Information Technology} \\
			\textit{University of Cincinnati}\\
			Cincinnati, Ohio, USA \\
			elsayeny@ucmail.uc.edu}
		\and
		\IEEEauthorblockN{Zag ElSayed}
		\IEEEauthorblockA{\textit{School of Information Technology} \\
			\textit{University of Cincinnati}\\
			Cincinnati, Ohio, USA \\
			elsayezs@ucmail.uc.edu}
		\and
		\IEEEauthorblockN{Said Varlioglu}
		\IEEEauthorblockA{\textit{School of Information Technology} \\
			\textit{University of Cincinnati}\\
			Cincinnati, Ohio, USA \\
			varlioms@mail.uc.edu}
	}
	
	\maketitle
	
	\thispagestyle{plain}
	\pagestyle{plain}
	
	\begin{abstract}

		Drug overdose is a pressing public health concern in the United States, resulting in a significant number of fatalities each year. In this study, we employ the Journey to Crime (JTC) framework borrowed from the field of environmental criminology to examine the association between drug sales locations and overdose death locations. The JTC framework draws on the concept of distance decay, positing that criminal activities are more likely to transpire in proximity to offenders' residential areas. In this research, our objective is to elucidate the trajectory of overdose victims to overdose locations, aiming to enhance the distribution of overdose services and interventions. To the best of our knowledge, no previous studies have applied the JTC framework to investigate drug overdose deaths. Consequently, our study represents a novel application of this framework to the realm of drug offenses. By scrutinizing data obtained from the Hamilton County, OH Coroners, and the Cincinnati Police Department, we endeavor to explore the plausible correlation between overdose deaths and drug sales locations. Our findings underscore the necessity of implementing a comprehensive strategy to curtail overdose deaths. This strategy should encompass various facets, including targeted efforts to reduce the accessibility of illicit drugs, the enhancement of responses to overdose incidents through a collaborative multidisciplinary approach, and the availability of data to inform evidence-based strategies and facilitate outcome evaluation. By shedding light on the relationship between drug sales locations and overdose death locations through the utilization of the JTC framework, this study contributes valuable insights to the field of drug overdose prevention. It emphasizes the significance of adopting multifaceted approaches to address this public health crisis effectively. Ultimately, our research aims to inform the development of evidence-based interventions and policies that can mitigate the occurrence and impact of drug overdoses in our communities.
	\end{abstract}
	
	\section{Introduction}
	\footnote{Preprint.  This paper will be presented at the 7th International Conference on Applied Cognitive Computing (ACC'23); Publisher: CPS; July 24-27, 2023; Las Vegas, Nevada, USA; https://american-cse.org/csce2023/conferences-ACC} Drug overdose has become a significant public health concern in the United States, with tens of thousands of people dying each year due to this cause\footnote{CDC WONDER. (2020). Wide-ranging online data for epidemiologic research (WONDER). Atlanta, GA: CDC, National Center for Health Statistic. Available at http://wonder.cdc.gov.}. Over the past decade, the number of drug overdose deaths has increased by 120\%, reaching nearly 70,630 per year by 2019\footnote{World Health Organization (2021). Opioid overdose. Retrieved Sep 2, 2021 from https://www.who.int/news-room/fact-sheets/detail/opioid-overdose}. In addition, during the COVID-19 pandemic\footnote{World Health Organization. (2021). Ibid.}, there was a 20\% increase in drug overdose deaths in 25 states and the District of Columbia\footnote{CDC. (2020a). Increase in Fatal Drug Overdoses Across the United States Driven by Synthetic Opioids Before and During the COVID-19 Pandemic. Retrieved Sep 3, 2021 from https://emergency.cdc.gov/han/2020/han00438.asp}. The World Health Organization recommends that naloxone administration can prevent opioid overdose deaths if given in time. Therefore, this study aims to understand the journey to overdose locations using the Journey to Crime (JTC) framework from environmental criminology to improve overdose service distribution and interventions. This framework is based on the distance decay and familiarity principles, which suggest that crimes are likely to occur closer to the offenders' residential areas and daily routes. The study will explore the relationship between drug sales locations and overdose deaths in Hamilton County, OH, using the JTC framework. This paper aims to introduce the JTC framework, its application to drug crimes and overdoses, the distance and buffer zones related to the framework, and the policy implications of the study. 
	
	\section{The Journey to Crime (JTC) Framework}
	The study of crime's spatial characteristics gained popularity in the 1960s and was later formalized as environmental criminology by Bratingham and Bratingham in 1981 \cite{wortley2008environmental}. The Journey to Crime framework is a subset of environmental criminology based on the concept of familiarity/distance decay. This theory suggests that offenders are more likely to commit crimes closer to their residences and that the chance of committing a crime decreases as the offender moves away from familiar locations such as their daily routes and home \cite{brantingham1993nodes} \cite{felson2003simple} \cite{wiles2000road}. Various studies support the distance/decay function for different crime types, including violent, property, and drug crimes \cite{santtila2007crime} \cite{block2009finding} \cite{levine2013journey}.
	
	Journey to crime suggests that offenders make rational choices, using the shortest path possible to commit a crime. This principle is related to Zipf's (1949) \cite{zipf1949human} "least effort principle." Offenders look for criminal opportunities that require minimal time and effort. This framework also implies a comfort-zone principle since offenders are more likely to commit crimes in familiar locations. Therefore, the shortest distance principle is related to the least effort and comfort-zone principles \cite{weisburd2006does}.
	Studies indicate that the average distance traveled to a crime location is about three miles \cite{sarangi2006spatial}\cite{snook2004individual} \cite{wiles2000road}. However, the distance varies based on different types of crimes and offender characteristics. For instance, violent criminals tend to travel shorter distances than property crime offenders \cite{bullock1955urban}\cite{groff2006exploring}\cite{reid2014uncovering}\cite{rossmo2016geographic}\cite{wiles2000road}\cite{white1932relation}. Additionally, white offenders travel longer distances compared to minority offenders \cite{ackerman2015far}\cite{groff2006exploring}\cite{phillips1980characteristics}. Moreover, older criminals \cite{gabor1984offender}\cite{rouwendal1994changes}\cite{warren1998crime} and male offenders \cite{gabor1984offender}\cite{groff2006exploring}\cite{pettiway1995copping} tend to travel further to commit crimes.
	Although the distance decay function suggests a linear relationship, studies suggest that offenders do not commit crimes within their immediate environment, known as buffer zones \cite{bernasco2005residential}\cite{brantingham1984patterns}. Studies have found weak to moderate evidence for the buffer zone idea of the journey to crime framework \cite{bernasco2005residential}\cite{block2007journey}\cite{groff2006exploring}.

	\vspace{1em} 
	\section{Journey to Crime and the Current Study}
	There is a limited amount of research available on the topic of the Journey to Crime (JTC) framework and drug offenders. However, based on the available studies, we do know that drug users tend to travel longer distances to areas with higher levels of deprivation where there is more availability of drugs \cite{forsyth1992geographical}\cite{johnson2016drug}\cite{johnson2013need}. In addition, studies suggest that drug offenders often travel longer distances to purchase drugs with fewer risks, which may be related to racial differences in drug markets \cite{donnelly2021opioids}\cite{johnson2020exploring}\cite{johnson2013need}\cite{levine2013journey}.
	Donnelly et al. (2021)\cite{donnelly2021opioids} conducted a study specifically focusing on Black-White differences in JTC and opioid possession offenses. Their research builds upon the social disorganization theory of Shaw and McKay (1942)\cite{shaw1942juvenile} and argues that concentrated disadvantage accumulates in inner cities where Black communities mostly reside. Therefore, Whites may travel longer distances from their original locations with higher economic standards to purchase opioids, as these deprived areas offer more opportunities for drug buyers.
	Previous studies also suggest that drug offenders tend to travel more for expensive drugs, such as heroin \cite{forsyth1992geographical}\cite{johnson2013need}.
	The current study aims to apply the JTC framework to drug overdose death locations to identify any patterns that could improve service distributions like naloxone. This study is the first of its kind to specifically examine the journey to drug overdose death locations, as previous studies have only focused on the residential distances of offenders to drug crimes.
	
	\section{Data}
	For this study, we utilized two primary data sources. The first dataset was provided by the Cincinnati Police Department and included a range of datasets, such as reported crimes (N=125,759), arrest data (N=83,786), and field interview reports (FIR) (N=291,443), spanning from January 1, 2013, to December 31, 2016. The arrest and incident data allowed us to identify drug-related crimes, while the field interview reports documented police contact and included the location of the contact.
	The second dataset we used came from the Hamilton County Coroner's office. This dataset contained information on overdose victims, including their names, date of birth, location of overdose death, residential address, demographic variables (sex, race, marital status), and cause of death for the year 2016 (N=195). This dataset allowed us to analyze overdose death locations and identify patterns that could potentially improve naloxone distribution.
	\section{Analytical Strategy}
	We will use arrest and suspect data to identify prolific drug sellers. The study will employ FIR data to determine drug sellers’ active locations in the city. In this way, we will have a more robust picture regarding the current drug sales locations in 2016. On the other hand, we will employ overdose data to see the characteristics of overdose death locations and distance to drug sales locations.
	This study will use graph-link theory to see the drug sales network. After identifying the illicit drug network, we will geocode FIR data to identify the main locations of prolific drug sellers, which will provide dynamic drug sales locations. For this purpose, we will have ArcMap 10.4 for geocoding and geo-analysis. The final dataset will be analyzed using negative binomial regression of R statistics since the dependent variable is overly dispersed, as seen in Table 1.
	
	\begin{table*}[t]
		\caption{Descriptive Statistics.}
		\small
		\begin{center}
			\begin{tabular}{lcccc}
				\textbf{Demographic Variables} & \textbf{N} & \textbf{Min} - \textbf{Max} & \% / \textbf{Avg.} & \textbf{s.d.} \\    
				\hline
				Gender (Male = 1) & 195 & 0 - 1 & 64.10\% & - \\
				Age & 195 & 19 - 83 & 41.76 & 12.1 \\
				Race (Black = 1) & 195 & 0 - 1 & 16.41\% & - \\
				Marital Status (Married = 1) & 195 & 0 - 1 & 15.90\% & - \\
				Education & 195 & 1 - 7 & 3.010 & 1.167 \\
				\hline    
				\textbf{Distance Variables} & & & & \\
				\hline 
				Journey to Overdose & 195 & 0 - 58.10 & 5.818 & 10.685 \\
				Overdose Location to Drug Sales Location & 195 & 0 - 5.43 & 0.756 & 1.27 \\
				\hline
			\end{tabular}
			\label{tab:descriptive-stats}
		\end{center}
	\end{table*}

	\section{Measures of Variables}
	\subsection{Dependent Variable}
	Journey to overdose (JTO) is the dependent variable of the study. We measured the distance of the overdose victims’ residents to overdose locations using Google Map; therefore, the distance reflects the actual travel distance in miles. Table 1 below shows that the range of the JTO is from 0 miles (at resident, n=91) to 58.1 miles (n=104). The average travel mile is 5.82 miles. Each overdose victim deviates from another victim by 10.69 miles which is the sign of overdispersion. 
	
	\subsection{Independent Variables}
	\textbf{Gender:} Descriptive statistics in Table 1 display that 64.1\% of the overdose victims are male. Past studies found that males travel more to commit crimes. We also hypothesized that males (coded as 1) compared to females (coded as 0) travel more to overdose.
	
	\textbf{Age:} Similarly, past studies suggest that older people travel more to commit crimes. Concurrent to this finding, we hypothesized that older people travel more to overdose. Age in the sample ranges from 19 to 83 years old, average 42 years old. 
	
	
	\begin{table*}[t]
		\caption{Source of Relationships for Social Network Analysis.}
		\small
		\begin{center}
			\begin{tabular}{lcccc}
				\hline
				& \textbf{First Degree (N)} & \textbf{Second Degree (N)} & \textbf{Third Degree (N)}  & \textbf{Total}\\
				\hline
				\textbf{Arrested together} & 3,025 & 1.472 & 574 & 5.071 \\
				\textbf{Field Interview Reports} & 15,033 & 5,974 & 2,018 & 23,025 \\
				\hline
				\textbf{Total} & 18,058 & 7,746 & 2,546 & 28,096 \\
				\hline
			\end{tabular}
			\label{tab:relationship_source}
		\end{center}
	\end{table*}
	
	\textbf{Race:} Many studies studied the role of race in JTC literature. The general finding suggests that whites travel more to commit a crime. A socio-economic explanation for this finding suggests that Blacks live in disadvantaged areas associated with high crime activity. In our study, 16.4\% of the sample is black. Parallel to the literature, we hypothesized that Blacks (coded as 1) compared to Whites (coded as 0) travel more to overdose.
	
	\textbf{Marital Status:} Almost 16\% of the sample’s marital status is married. We hypothesized that married people travel more because they are more likely to live far from drug sales locations.
	
	\textbf{Education:} Education is generally associated with the socio-economic status of people. Therefore, we hypothesized that higher education increases travel distance. The breakdown of the sample for education is as follow: 8th grade or less (n=12), 9th through 12th grade-no diploma (n=53), high school graduate or GED completed (n=78), some college but no degree (n=24), associate degree (n=13), bachelor's degree (n=6).
	
	\textbf{Distance from Overdose Location to Drug Sales Location:} We generated this variable by taking the distance from the overdose location to the identified drug sales locations discussed above. We hypothesized that victims living far away from the known overdose locations would travel more and overdose closer to the drug sales locations. Putting it differently, as overdose travel distance increases, victims are more likely to overdose near overdose drug sales locations to relieve the addiction pressure sooner. The average distance to drug sales locations is 0.76 miles, which is fairly short in the sample.

	\begin{figure*}[ht] 
		\centering
		\includegraphics[width=12cm,height=5cm]{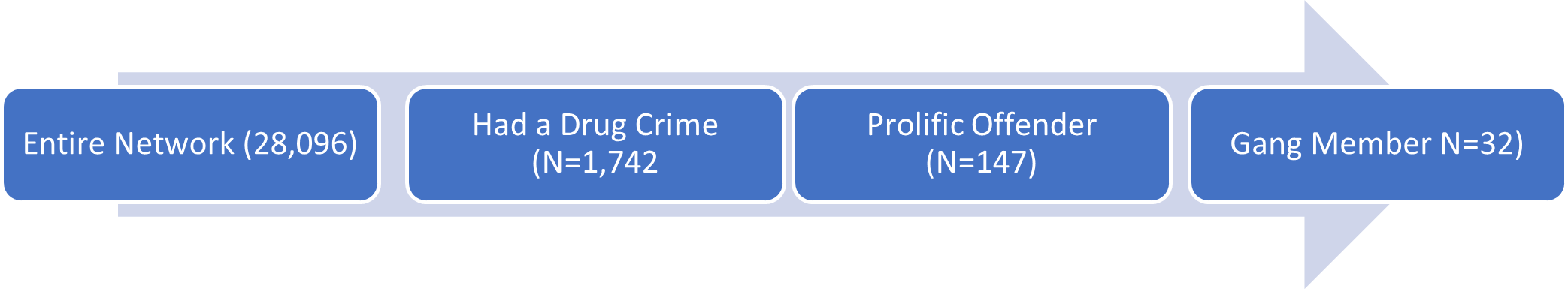}
		\vspace{-0.5cm} 
		\caption{Understanding Illicit Drug Network.}
		\label{fig1}
	\end{figure*}
	
	\section{Results}
	To identify illicit drug networks, we first employed link analysis to determine the co-offending networks using arrest and FIR (Field Interview Report) data. Link analysis is a powerful tool commonly used in criminology to identify and analyze relationships between individuals involved in criminal activities \cite{papachristos2012social}. Recent studies have shown that associations within co-offending networks provide better explanations of criminal behavior than factors such as gender, race, or gang affiliation \cite{papachristos2012social}.
	
	In our study, we adopted the assumption that if two or more individuals were arrested together or stopped by the police for a field interview, it suggested a potential connection between them. By analyzing the patterns of co-offending, we aimed to uncover the hidden structure of drug-related criminal networks and explore their implications for drug sales locations and overdose deaths.
	
	Link analysis has proven to be a valuable approach in understanding the dynamics of criminal networks. It allows researchers to identify key players, detect patterns of collaboration, and gain insights into the overall structure and functioning of these networks \cite{kleemans2011criminal}. By applying this methodology to our research on drug offenses and overdose deaths, we sought to reveal the interconnections between individuals involved in drug-related activities and the locations where these activities take place.
	
	By utilizing arrest and FIR data, we aimed to construct a comprehensive network of individuals involved in drug offenses. This network analysis approach allowed us to examine the relationships and interactions among offenders, potentially leading us to identify central figures or key hubs within the drug distribution network. Furthermore, understanding the spatial distribution of drug sales locations and their proximity to overdose death locations can provide valuable insights for developing targeted interventions and prevention strategies.
	
	\begin{figure*}[t]
		\centerline{\includegraphics[width=18.5cm,height=14cm]{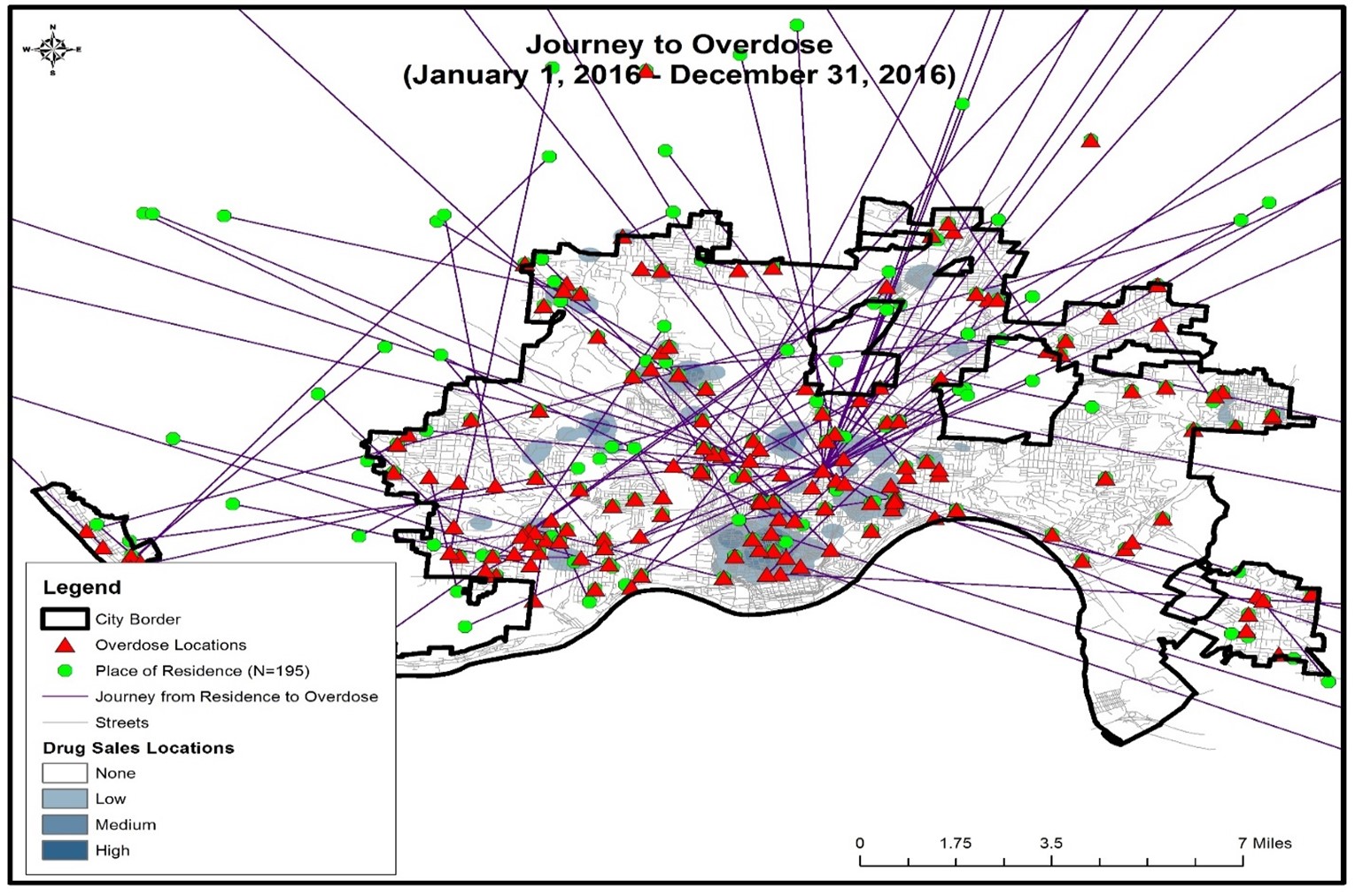}}	
		\caption{Journey to Overdose.}
		\label{fig2}
	\end{figure*}
	
	Given this context, analysis of co-offending networks in Table 2 revealed that 28,096 individuals are nested in the network. The most prominent type of relationship is field interview reports. However, the current analysis is too broad and fails to inform decision-makers about the illicit drug network since the co-offending network includes all crimes. Therefore, as shown in Figure 1, we dropped the relationships/edges if they did not include any drug-related crimes, and our sample size significantly dropped to 1,742 relationships following this process. In addition to this, we only had individuals in the sample if they were prolific offenders (e.g., prior record of violent crimes) and arrested more than once for drug-related offenses.

	After this filtering process, the entire sample is reduced to 147 prolific offenders who had criminal records for repeatedly selling drugs. Further analyses showed that 32 out of the 147 prolific offenders are gang members. The remaining majority of prolific offenders are connected to gang members at the first, second, and third-degree levels of their friendship network. This analysis suggests that only a handful of prolific-gang-affiliated criminals are responsible for the drug market in Cincinnati. The current findings are also aligned with gang literature that posits majority of gangs sell drugs to make money \cite{engel2013reducing}\cite{national2012}.

	\begin{table*}[t]
		\caption{Negative Binomial Results of Journey to Overdose.}
		\small
		\begin{center}
			\begin{tabular}{lcccc}
				\hline
				Variable & b & s.e. & z value & Pr($>|z|$) \\
				\hline
				Intercept & 3.601 & 0.385 & 9.355 & $<$ 0.000 *** \\
				Overdose Location to Drug Sales Location & -0.245 & 0.067 & -3.647 & 0.0003 *** \\
				Education & 0.304 & 0.068 & 4.442 & $<$ 0.000 *** \\
				Gender & -0.017 & 0.166 & -0.101 & 0.919 \\
				Age & -0.064 & 0.007 & -9.283 & $<$ 0.000 *** \\
				Marital Status & -0.578 & 0.225 & -2.568 & 0.010 * \\
				Race & -0.589 & 0.224 & -2.637 & 0.008 ** \\
				\hline
			\end{tabular}
			\label{tab:relationship_source1}
		\end{center}
	\end{table*}

	Following the identification of illicit drug networks, the analysis of spatial patterns becomes crucial in understanding the dynamics of drug-related activities. Map analysis provides valuable insights into the geographic distribution of drug sales locations and their relationship with overdose deaths. Several studies have highlighted the significance of map analysis in exploring the spatial aspects of criminal behavior and its implications for public health interventions \cite{weisburd2008putting} ,\cite{ma2019spatial}.
	
	In this study, we emplyed map analysis techniques to examine the spatial patterns of drug sales locations and their association with overdose deaths. By utilizing kernel density mapping, we created a visualization (Figure 2) that displays the frequent police contact locations of prolific drug sellers in shaded blue areas. This representation allows for a clear identification of areas with a high concentration of drug-related activities.
	
	Furthermore, we geocoded overdose data using both the overdose death locations and the resident addresses of the victims. The green-colored circles on the map indicate the resident addresses of the overdose victims (N=195), while the red-colored triangles represent the locations of overdose deaths. The journey from the victims' residences to the overdose locations is represented by straight lines.
	
	Upon examining the map, two notable inferences can be drawn. First, victims residing at greater distances from drug sales locations tend to travel more to reach the overdose sites. This suggests that access to drug markets plays a crucial role in shaping the spatial distribution of overdose incidents. Second, there is a significant overlap between the locations of overdose deaths and drug sales areas, indicating a potential association between the availability of illicit drugs and the occurrence of fatal overdoses.
	
	To gain a deeper understanding of these observed spatial patterns and draw statistical inferences, we conducted regression analysis. Prior to the analysis, we performed tests for collinearity and found no evidence of high correlation among the bivariate variables. This ensures the validity of our regression analysis and allows us to examine the relationship between various factors, such as distance to drug sales locations, overdose deaths, and other relevant variables.

	Negative binomial regression in Table 3 suggests that, except for gender, all variables are significantly related to the outcome variable. More specifically, education is positively associated with travel distance, suggesting that higher education increases overdose journeys. The education coefficient suggests that increasing one education level between the ordinal categories of the variable results in 35.56\% more travel distance (exp(.304)=1.3553).  
	On the other hand, age, marital status, and race are negatively related to the dependent variable. The standard deviation of the age variable in the sample is 12.1. Therefore, one standard deviation in age results in 53.9\% less travel distance. Restating it with plain language, a 42 year old person travels 53.9
	
	Findings also suggest that married people travel 43.9\% less distance compared to singles. This finding is against our hypothesis because we envisioned that married people are more likely to live away from inner cities. On the other hand, as aligned with the expected direction, Blacks travel 44.5\% less than whites to overdose. 
	The distance of the overdose location to drug sales locations is negatively and significantly associated with the journey to overdose. That means, as the distance between overdose locations and drug sales locations increases, the journey to overdose decreases. This finding suggests that victims who travel more to overdose locations tend to overdose closer to the drug sales locations, as hypothesized in the study. Mathematical interpretation suggests that for each standard deviation (1.27 miles), a change in the “overdose location to drug sales locations” variable yields 26.75\% less travel distance. 
	
	\section{Discussion and Conclusion}
	
	The findings of this study provide valuable insights into the relationship between drug sales locations and overdose deaths in Hamilton County, OH, using the Journey to Crime (JTC) framework from environmental criminology. By applying the JTC framework, this study aimed to improve overdose service distribution and interventions by understanding the journey to overdose locations.
	
	The JTC framework, based on the familiarity/distance decay concept, suggests that offenders are more likely to commit crimes closer to their residences and familiar locations. This study found that the average travel distance to overdose locations was 5.82 miles, with a range from 0 to 58.1 miles. These findings align with previous research on crime distances, which indicate that offenders tend to commit crimes within a relatively short distance from their homes. However, it is important to note that distance decay function does not imply that crimes occur within immediate environments, as buffer zones play a role in crime occurrence.
	
	In the context of drug crimes and overdoses, previous studies have shown that drug users tend to travel longer distances to areas with higher levels of deprivation and greater drug availability. This study builds upon these findings and examines the journey to overdose death locations, which is a novel contribution to the literature. The results revealed that victims living far distances from drug sales locations tended to travel more to overdose. This suggests that individuals who reside farther away from known drug sales locations may be willing to travel longer distances to obtain drugs, potentially indicating a higher level of desperation or limited access to drugs in their immediate vicinity.
	
	Furthermore, the study explored the demographic factors that may influence the journey to overdose. The analysis revealed that gender, age, race, marital status, and education were significant factors associated with the distance traveled to overdose locations. Males and older individuals tended to travel more to overdose, which aligns with previous research indicating that males and older offenders travel longer distances to commit crimes. The findings related to race showed that Whites tended to travel more to commit crimes, which may be linked to socio-economic factors and the distribution of drug markets. Additionally, marital status and education were associated with the distance traveled, with married individuals and those with higher education levels traveling more. These findings provide valuable insights into the socio-demographic characteristics of individuals who travel longer distances to overdose locations.
	
	The analysis of the illicit drug network and the identification of prolific drug sellers revealed that a small number of individuals, including gang members, were responsible for the drug market in Cincinnati. This finding is consistent with previous research highlighting the role of gangs in drug sales. Understanding the dynamics of the drug market and the locations of prolific drug sellers is crucial for effective interventions and service distribution, such as naloxone distribution, to target high-risk areas.
	
	The policy implications of this study are significant. By utilizing the JTC framework and understanding the journey to overdose locations, interventions can be designed and targeted more effectively. The findings suggest the need for targeted naloxone distribution strategies, with a focus on areas with higher drug sales activity and where individuals may be traveling longer distances to obtain drugs. This can help ensure that naloxone is readily available in the locations where it is most needed to prevent overdose deaths. Additionally, the identification of prolific drug sellers and the understanding of the illicit drug network can inform law enforcement efforts and strategies to disrupt the drug market.
	
	\section{Conclusion}
	The analysis of the illicit drug network and the identification of prolific drug sellers emphasized the concentration of drug sales in the hands of a few individuals, including gang members. This understanding is crucial for implementing effective law enforcement strategies to disrupt the drug market and mitigate drug-related harms.
	
	The policy implications of this study are significant for addressing the overdose crisis. By leveraging the insights provided by the JTC framework and understanding the journey to overdose locations, targeted interventions can be developed and implemented. Specifically, the findings suggest the need for focused naloxone distribution strategies, ensuring that this life-saving medication is readily available in areas with high drug sales activity and where individuals may travel longer distances to obtain drugs. This targeted approach can improve the effectiveness of naloxone distribution efforts, which may then reduce the number of overdose deaths.
	
	In addition, the findings related to demographic factors highlight the importance of considering socio-demographic characteristics in designing interventions. Therefore, tailoring prevention and harm reduction programs to address the specific needs and circumstances/populations (e.g., males, older individuals, and those with higher education levels) can enhance the impact of these initiatives.
	
	There are certain limitations in the study. The current study focused on Hamilton County, OH, and may not be directly generalizable to other regions. Neverthelss, the methodology and insights derived from this research can serve as a valuable foundation for future studies in different contexts. In addition, future studies might include more cases to better understand the relationship between drug sales locations and overdose deaths.
	
	In conclusion, this study sheds light on the journey to overdose locations and its association with drug sales locations in Hamilton County, OH. By applying the JTC framework, the study deepens our understanding of the dynamics of drug-related harms and provides valuable insights for the development of targeted interventions and policies to address the overdose crisis. Future research building upon these findings can contribute to evidence-based strategies that effectively combat drug-related harms and save lives.
	
	\bibliographystyle{ieeetr}
	\bibliography{references}
	
\end{document}